\documentclass[prb,twocolumn,floatfix,superscriptaddress]{revtex4-1}

\usepackage{float}
\usepackage[caption = false]{subfig}
\usepackage[final]{graphicx}

\usepackage{dcolumn}
\usepackage{bm}
\usepackage{amssymb,amsmath}
\usepackage{color}
\usepackage{epstopdf}
\usepackage{bibunits}

\begin{document}

\title{Fluctuation-induced spin nematic order in magnetic charge-ice}

\author{A. Hemmatzade}
\affiliation{Laboratory for Neutron Scattering and Imaging, Paul Scherrer Institut, CH-5232 Villigen PSI, Switzerland}
\author{K. Essafi}
\affiliation{Laboratory for Theoretical and Computational Physics, Paul Scherrer Institut, CH-5232 Villigen PSI, Switzerland}
\author{M. Taillefumier}
\affiliation{CSC-Swiss National Supercomputing Centre, ETHZ, Lugano CH-6900, Switzerland}	
\author{M. M\"uller}
\affiliation{Laboratory for Theoretical and Computational Physics, Paul Scherrer Institut, CH-5232 Villigen PSI, Switzerland}
\author{T. Fennell}
\email{tom.fennell@psi.ch}
\affiliation{Laboratory for Neutron Scattering and Imaging, Paul Scherrer Institut, CH-5232 Villigen PSI, Switzerland}
\author{P. M. Derlet}
\email{peter.derlet@psi.ch} 
\affiliation{Laboratory for Theoretical and Computational Physics, Paul Scherrer Institut, CH-5232 Villigen PSI, Switzerland}
\begin{abstract}

Disorder in materials may be used to tune their functionalities, but much more strikingly, its presence can entail entirely new behavior. This happens in charge-ice where structural disorder is not weak and local, but strong and long-range correlated. Here, two cations of different charge occupy a pyrochlore lattice, arranging themselves such that all tetrahedra host two cations of each type. The ensuing correlated disorder is characterized by randomly packed loops of a single cation-type. If the cations are magnetic and interact antiferromagnetically, a new type of magnet with strong interactions along the loops, but frustrated interactions between loops, emerges. This results in an ensemble of intertwined Heisenberg spin chains that form an algebraic spin liquid at intermediate temperatures. At lower temperatures, we find these non-local degrees of freedom undergo a discontinuous transition to a spin nematic. While this phase does not break time reversal symmetry, its spin symmetry is reduced resulting in a dramatically slower spin relaxation. The transition is sensitive to the statistics of the cation loops, providing both a direct thermodynamic signature of otherwise elusive structural information and a structural route to engineering nematic phase stability.

\end{abstract}
\maketitle

\begin{bibunit}

The discrete translational symmetry of the crystalline state underlies the utility of many functional materials. Introducing random disorder into crystalline materials can play a crucial role in modifying their static and dynamical properties to obtain new or improved functionality, for example by producing pinning centers in superconductors~\cite{Blatter1994} or tuning transition temperatures in multi-ferroics~\cite{Morin2016,Scaramucci2020}. Recently it has been suggested that a kind of correlated disorder based on tiling high symmetry lattices with low symmetry motifs may be a route to novel functionalities via the interplay of the disorder with crystal properties such as lattice dynamics or electronic conduction~\cite{Overy2016}. Magnetism is another material property that may be controlled by disorder. Usually uncorrelated variations of exchange strength or coordination (via uncorrelated doping of magnetic ions) are expected to produce spin glasses~\cite{Binder1986}. Here, we show that more correlated types of structural disorder may result in distinct equilibrium and out-of-equilibrium properties.

\begin{figure*}
\includegraphics[width=0.8\linewidth,trim=0cm 0cm 0cm 0cm,clip]{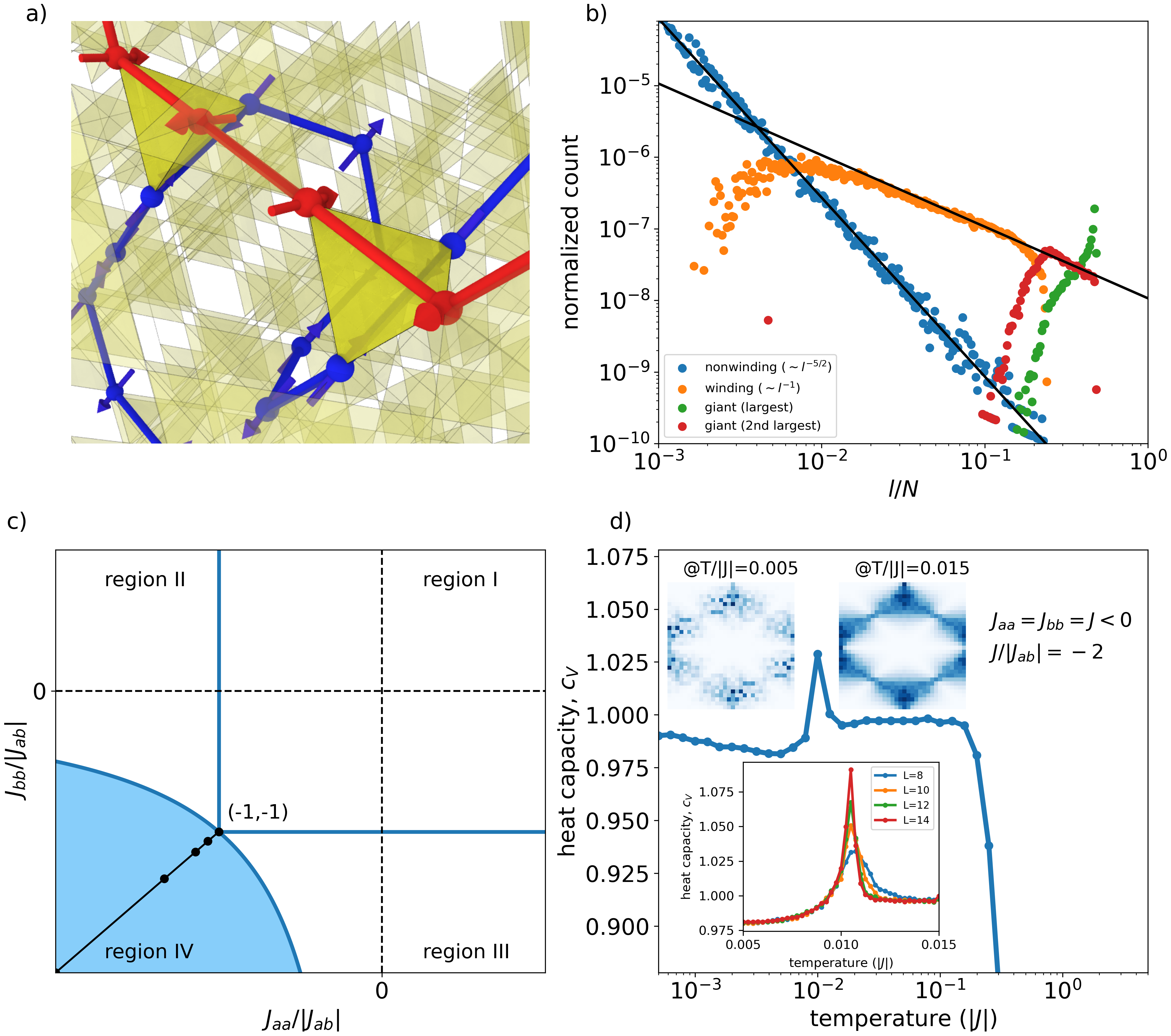}
\caption{{\bf Charge-ice correlated exchange disorder and its low temperature magnetic properties.} a) The charge-ice structure of two types of cations, $a$ and $b$, distributed on the pyrochlore lattice is characterized by an ensemble of closed loops of the same atom species. Two loops of different species (blue and red) may share a number of tetrahedra. b) Sample-averaged normalized histogram of loop lengths in which non-winding and winding are distinguished, as well as the two largest (giant) loops. The power-law exponents are well understood from the perspective of diffusion~\cite{Jaubert2011}. The normalized distributions of loop length $l$  are plotted as a function of $l/N$ where $N=16L^{3}$ with $L=20$. c) Magnetic charge-ice is characterized by the nearest neighbour exchange constants $J_{aa}$, $J_{bb}$, and $J_{ab}$ between moments associated with species a and b. The ground-state phase diagram~\cite{Banks2012} is found to be rich in structure with regions I-III hosting long-range ordered phases. The boundary to region IV ($J_{aa}J_{bb}=J_{ab}^{2}$) hosts a classical Heisenberg pyrochlore AFM at $(-1,-1)$. Within IV the ground state is a (less degenerate) classical spin liquid with perfect AFM order on each loop but no correlations between loops. d) At low temperature, the heat capacity of region IV reveals a transition toward a phase in which the giant and other large loops are collinear. 
This nematic transition is reflected in the structure factor (upper inset) which has a smooth pinch point structure above the transition and becomes patchy below it.}
\label{fig1}
\end{figure*}

An ice rule is a well known way to introduce such correlated disorder. Here, a simple constraint on the local configuration of binary degrees of freedom allows the construction of an extensively degenerate manifold of states, in which the correlation among local degrees of freedom decays not exponentially, but by a $1/r^3$ (dipolar) power-law~\cite{Henley2010}. Such rules have become common currency for describing spin configurations in geometrically frustrated magnetic materials and arrays of nano-magnets, notably spin ice~\cite{Harris1997}, quantum spin ice~\cite{Gingras2014}, and artificial spin ice~\cite{Skjaevo2020,Schiffer2021}. The equivalence of spins and charges on this lattice was first noted by Anderson in an investigation of the Vewey transition in magnetite~\cite{Anderson1956}, where it was pointed out that cations would obey the charge-ice rule which requires that each tetrahedron is occupied by two cations of each type. If the different cations carry magnetic moments one obtains a model of magnetic charge-ice, which is directly relevant to pyrochlores of the type AMM'F$_6$ (for example CsCrNiF$_{6}$, see Ref.~\cite{Fennell2019}). In pyrochlores such as R$_2$MM'O$_7$ and AA'M$_2$F$_7$, see respective Refs.~\cite{Simonet2023,Plumb2019}, the charge-ice formed amongst the non-magnetic spectator ions may introduce more subtly correlated bond disorder amongst the magnetic atoms. More generally, geometric frustration of charge order causes correlated distributions of species and thus of the interactions among their associated degrees of freedom, resulting in specific material properties that reflect the correlated nature of the underlying disordered structure.

{\bf Strong and correlated magnetic exchange disorder via a charge-ice.} Here we consider how an exchange network templated by a charge-ice cation configuration affects the low temperature properties of a classical Heisenberg spin system. In a simple model of magnetic charge-ice~\cite{Banks2012}, two types of magnetic atoms $a$ and $b$ populate the pyrochlore lattice according to the 2:2 charge-ice rule resulting in correlated site disorder characterised by a randomly packed set of single cation-type nearest-neighbour connected loops of even length (Fig.~\ref{fig1}a). Fig.~\ref{fig1}b shows the loop size distribution in which we distinguish four classes of loops: those that are non-winding or winding with respect to the periodic boundaries of the system, and, additionally, the largest and second-largest winding loop. For a given charge-ice realisation these latter two will be of different chemical type and we refer to them as giant loops. The fraction of sites occupied by the four loop classes tends (with increased sampling and system size) to $f_{\mathrm{nw}}=0.06$ for non-winding, $f_{\mathrm{w}}=0.22$ for winding, $f_{2}=0.31$ and $f_{1}=0.41$ for the second and largest (giant) loops respectively, in agreement with Ref.~\cite{Jaubert2011}.

We describe the magnetic structure by unit-length classical Heisenberg spins on the sites, that are connected by the nearest neighbour exchange constants $J_{\mathrm{aa}}$, $J_{\mathrm{bb}}$, and $J_{\mathrm{ab}}$, into which we absorb the size of the different cation magnetic moments. The resulting Heisenberg Hamiltonian displays correlated bond disorder that derives from the spatial structure of the cation loops. Banks and Bramwell~\cite{Banks2012} identified four regions of the ground-state phase diagram for this model, as shown in Fig.~\ref{fig1}c. We focus on region IV, where $J_{aa}J_{bb}>J_{ab}^2$, with $J_{aa}$ and $J_{bb}$ both promoting intra-species antiferromagnetic (AFM) alignment, so that the zero temperature ground states have perfect AFM arrangements on each loop, but are degenerate with respect to the orientation of the AFM alignment axis (the N\'{e}el vector) of any loop due to the inter-loop couplings $J_{ab}$ being perfectly frustrated. In the work of Ref.~\cite{Banks2012}, Monte Carlo simulations  at $T/|J|<0.12$ for regime IV ($J_{aa}=J_{bb}=J$ with $J/|J_{ab}|=-1.2$) revealed a pinch-point-like structure factor and a vanishing Edwards-Anderson parameter down to $T/|J|=0.012$ (indicating no spin freezing), suggesting an algebraic spin liquid.

\begin{figure*}
\includegraphics[width=0.8\linewidth,trim=0cm 0cm 0cm 0cm,clip]{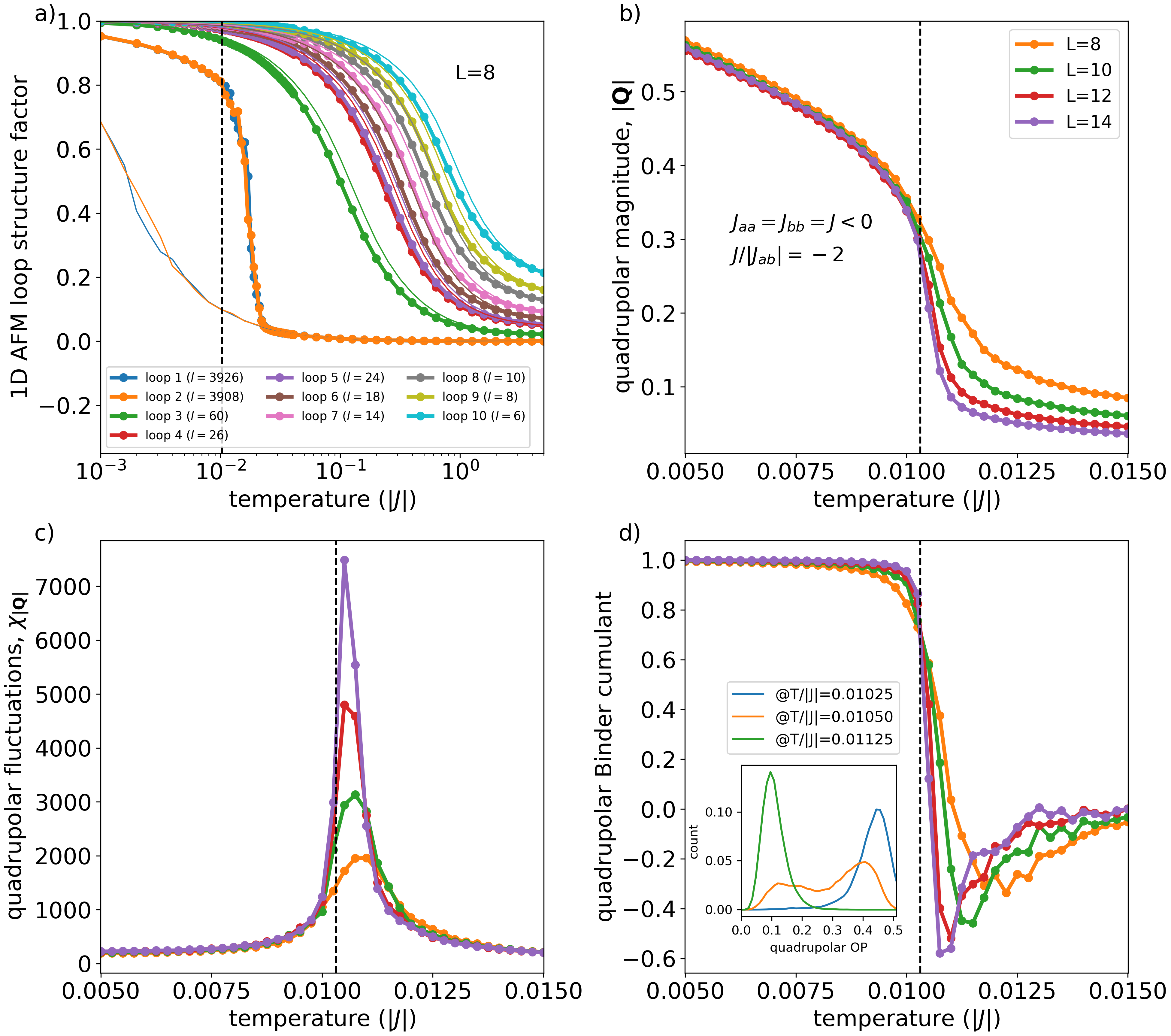}
\caption{{\bf A first order nematic phase transition.} a) The one dimensional AFM loop structure factor progressively increases as the temperature is reduced. Above $T_{0}$ all loops exhibit 1D domain wall activity and thus short range order at a finite temperature characteristic of finite Heisenberg spin-chains (thin lines represent the $J_{\mathrm{ab}}=0$ case of non-interacting loops). However, around $T_{0}$, the giant loops show a discontinuous jump in their AFM order as their N\'{e}el vectors align. Such alignment is diagnosed by the bulk quadrupolar order parameter $\mathbf{Q}$, b), whose average magnitude, $\langle|\mathbf{Q}|\rangle/N$ ($|\mathbf{Q}|^2\equiv {\rm Tr}[\mathbf{Q}^2]$), jumps at $T_{0}$ concomitant with strong fluctuations c) defined through the generalized susceptibility $\chi_{|\mathbf{Q}|}=(\langle|\mathbf{Q}^{2}|\rangle-\langle|\mathbf{Q}|^{2}\rangle)/(TN)$. The first order nature of this phase transition is revealed by the generalized Binder cumulant~\cite{Binder1992} which becomes negative close to $T_{0}$ indicating a bi-modal order parameter distribution (histograms of $|\mathbf{Q}|$ below, close-to, and above $T_{0}$ are shown in the inset). In all panels, an estimate of the infinite size $T_{0}/J$ equal to 0.0103 is indicated by the vertical dashed lines derived from  the Binder cumulant crossing seen in panel d).}
\label{fig2}
\end{figure*}

\begin{figure*}
\includegraphics[width=0.9\linewidth,trim=0cm 0cm 0cm 0cm,clip]{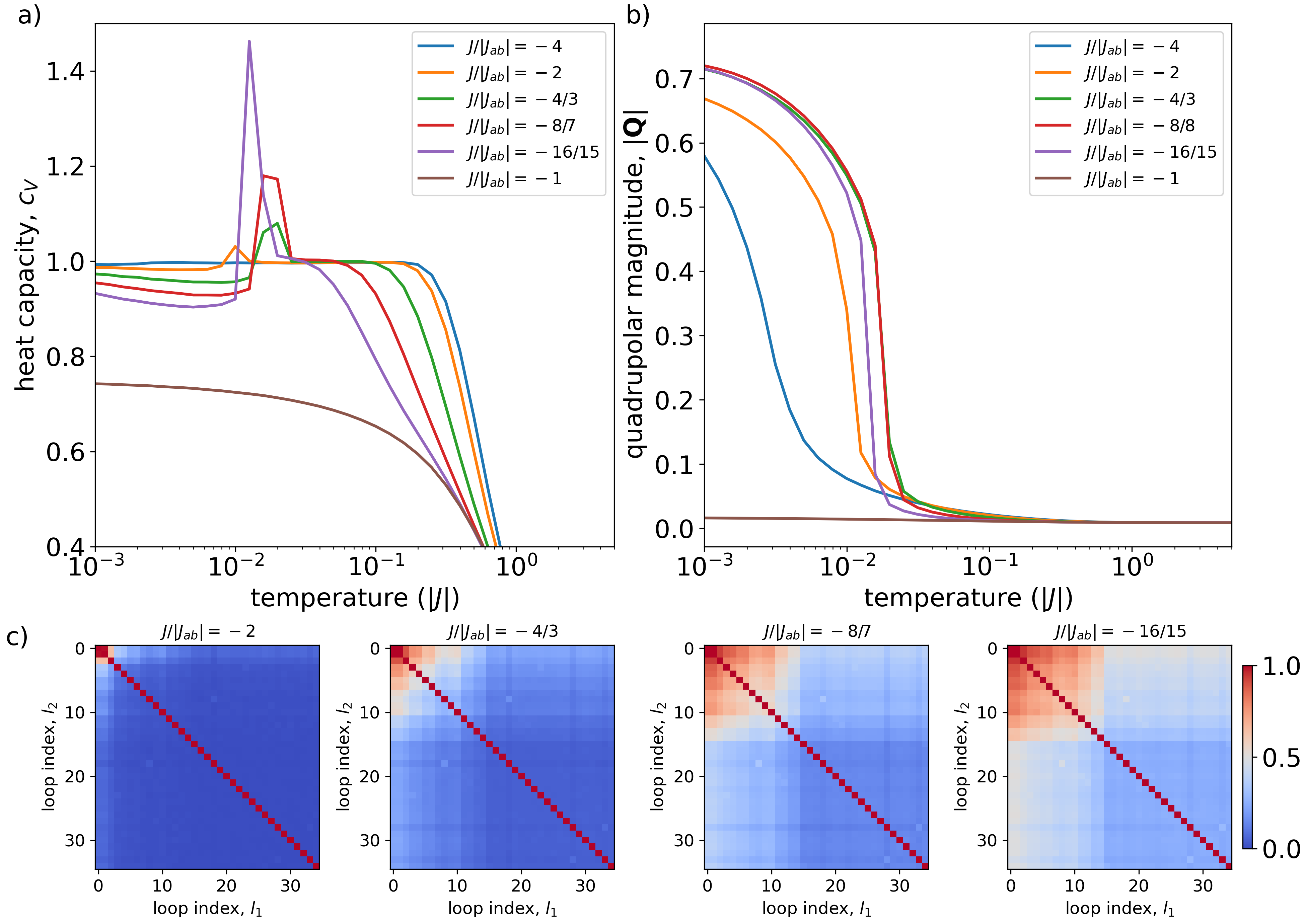}
\caption{{\bf Ordering depends on loop length and coupling strength.} Thermodynamic quantities of systems with different $J/|J_{ab}|$ within region IV and at its boundary, showing the a) heat capacity and b) average quadrupolar magnitude $|\mathbf Q|$ as a function of temperature. Except for the PHAFM ($J/|J_{ab}|=-1$), there is a phase transition whose $T_{0}$ is controlled by $J/|J_{ab}|$. Data is obtained using an $L=8$ sample with fixed charge-ice structure, containing loops of lengths 3926, 3908, 60, 26, 24, 18 (2$\times$), 14 (4$\times$), 10 (2$\times$), 8 (2$\times$), 6 (20$\times$). c) Loop-loop quadrupolar correlation matrix, $\langle{\rm Tr}[\mathbf{Q}_{l_1}\mathbf{Q}_{l_2}]\rangle/(\langle|\mathbf{Q}_{l_{1}}|^{2}\rangle\langle|\mathbf{Q}_{l_{2}}|^{2}\rangle)^{1/2}$, at the lowest $T/|J|=0.001$ for different loop-couplings. As $J/|J_{ab}|$ approaches -1 also smaller loops align increasingly and participate in the phase transition.}
\label{fig3}
\end{figure*}

{\bf Low temperature spin nematic order.} Focusing on the case $J_{aa}=J_{bb}=J<0$ with $J/|J_{ab}|=-2$ (region IV), our Monte Carlo heat bath algorithm~\cite{Miyatake1986} simulations show that on cooling ($0.125\lesssim T/|J|\lesssim1.5$) the system evolves from the paramagnetic state into a low temperature state with an energy per site approaching that of the expected ground state value, a strongly suppressed magnetization, and a well developed plateau in the heat capacity reflecting the low temperature behaviour of a classical Heisenberg system (Fig.~\ref{fig1}d). At our lowest temperatures, the heat capacity $c_V$ has a value just below unity, indicating a significantly more constrained system than the pyrochlore Heisenberg AFM (PHAFM), which asymptotes to $c_V=3/4$ (see Ref.~\cite{Moessner1998} and Fig.~\ref{fig3}a). A small peak in $c_V$ at $T_{0}/|J|\approx0.01$ (Fig.~\ref{fig1}d), which sharpens with increasing system size, suggests a previously unnoticed phase transition, whose nature we now elucidate.

The static structure factor (Fig.~\ref{fig1}d, inset panels), taken as  $S_{\mathbf{k}}=\left|N^{-1}\sum_{i=1}^{N}\mathbf{s}_{i}\exp(i\mathbf{k}\cdot\mathbf{r_{i}})\right|^{2}$ and thermally averaged over statistically independent spin configurations for a {\it single} charge-ice realisation, shows the distinctive diffuse scattering and pinch-points associated with dipolar spin correlations on the pyrochlore lattice above $T_{0}$. Below $T_{0}$ this pinch-point structure becomes patchy, like that of similarly sized individual charge-ice ground states~\cite{Banks2012}. The pinch-point structure arises because all spins that share the same loop are AFM correlated, which implies power-law spin correlations. This is not unlike the case of the PHAFM where the ground state manifold consists of {\em all} possible AFM close-packed loop realizations combining to give a smooth diffuse scattering profile at these system sizes. The patchiness is therefore due to a restricted sub-set of the full PHAFM manifold, which does not self-average at our finite system size. It can be removed by averaging over many charge-ice realizations or by considering larger system sizes.

To reveal the structure of the low $T$ phase we investigate the 1D AFM structure factor of the $i$th loop: $S^{\mathrm{loop}}_{\mathrm{AFM},i}=\left|l_{i}^{-1}\sum_{n=1}^{l_{i}}(-1)^{n}\mathbf{s}_{n}\right|^{2}$ where $l_{i}$ is its length. When $S^{\mathrm{loop}}_{\mathrm{AFM},i}=1$, the loop has complete AFM order, while the sign of its N\'{e}el vector may still fluctuate. Fig.~\ref{fig2}a plots $S^{\mathrm{loop}}_{\mathrm{AFM},l}$ for loops of various sizes as a function of temperature. Generally, they develop smoothly as the temperature is reduced, but for the two giant loops, the structure factor jumps up abruptly at a temperature $T_{0}$. As a reference, data are also shown for the case of non-interacting loops ($J_{\mathrm{ab}}=0$). Above $T_{0}$, loops of all sizes in the full system behave similarly as non-interacting loops, for which the thermal properties are known analytically~\cite{Fisher1964,Joyce1967}. Indeed, rescaling the temperature axis of the $J_{\mathrm{ab}}=0$ data by the factor $0.8$ results in almost perfect overlap with the $J_{\mathrm{ab}}\ne0$ data for $T>T_{0}$, suggesting the full system is well described in this temperature regime by an ensemble of non-interacting spin-chains with the renormalized coupling $\sim0.8J$. 

Inspection of the low temperature spin configurations reveals that for $T<T_{0}$ the N\'{e}el vectors of the two giant loops align collinearly, motivating the use of the bulk quadrupolar or nematic order parameter~\cite{Shannon2010}: $\mathbf{Q}=\sum_{i=1}^{N}\mathbf{Q}_{i}$, where $\mathbf{Q}$ is a traceless symmetric tensor with components $Q_{i}^{\mu\nu}= s_i^\mu s_i^\nu-\delta^{\mu\nu}/3$. A non-zero $\langle \mathbf{Q}\rangle$ signals breaking of rotational symmetry, but not necessarily of time reversal symmetry, being invariant under spin reversal (of entire loops). Figs.~\ref{fig2}b and c plot the average of the magnitude of the quadrupolar order parameter and a measure of its fluctuations near the transition, indicating a rapid turn-on of quadrupolar order that sharpens with increasing system size. Since $\mathbf{Q}$ and $-\mathbf{Q}$ describe qualitatively different spin structures the Landau free energy does not need to be invariant under a sign change of $\mathbf{Q}$ and will generally contain a cubic term ${\rm Tr}(\mathbf{Q}^3)$, ruling out a continuous phase transition. This is confirmed by the generalised Binder cumulant~\cite{Binder1992} for $\mathbf{Q}$, which becomes increasingly negative just above $T_{0}$ with increasing system size (Fig.~\ref{fig2}d), due to a bi-modal distribution of the order-parameter magnitude reflecting phase coexistence at $T_{0}$ (see inset in Fig.~\ref{fig2}d).

{\bf Role of loop lengths and loop-loop coupling.} Figs.~\ref{fig3}a-b display the heat-capacity and quadrupolar order parameter for a range of  $J/|J_{ab}|$ values within region IV. Also shown is $J/|J_{ab}|=-1$, which is the less constrained PHAFM and does not exhibit the nematic transition. Both observables show that $T_{0}$ increases as $J/|J_{ab}|$ increases, reaching a maximum around $J/|J_{ab}|=-8/7$ and then decreases indicating non-monotonic behavior very close to the PHAFM boundary. The magnitude of $|\mathbf Q|$ in the sub-$T_{0}$ temperature regime also increases,  indicating that a growing fraction of the sample nematically aligns. Defining $\mathbf{Q}_{l}$ as the quadrupolar order parameter of the $l$th loop, this trend is reflected in the loop-loop quadrupolar correlation function shown in Fig.~\ref{fig3}c. For $J/|J_{ab}|=-2$ the two giant loops dominate the transition and only their N\'{e}el vectors become well aligned. However, as $J/|J_{ab}|$ approaches -1, smaller and smaller loops take part in the alignment and contribute to the bulk quadrupolar order parameter. This trend saturates around $J/|J_{ab}|=-8/7$, where the smallest loops still remain only weakly aligned. 

These results might suggest the giant loops are the essential ingredient for the transition to occur. This is not the case, since breaking up the giant loops through a modified charge-ice algorithm (Fig.~\ref{figsm1}) or using open boundary conditions (not shown) has little effect on the transition as long as sufficiently large loops remain present. The observation that for a given $J/|J_{ab}|$, sufficiently small loops do not order, suggests that an ordered charge-ice structure, consisting of $4L^{2}$ linear loops of length $4L$ (system I, see Methods) should not nematically order for a small enough $L$. Indeed, for $L=8$, order is absent for moderate $J/|J_{ab}|=-2$, and only sets in for $J/|J_{ab}|\gtrsim -4/3$ (Fig.~\ref{figsm2}).

{\bf Order-by-disorder and symmetry reduction from  Heisenberg to Ising loops.} Insight into why the nematic structure is selected can be gained from the low temperature thermal properties of a single tetrahedron. Expanding the corresponding Hamiltonian to quadratic order with respect to transverse fluctuations around a ground state configuration defined by the angle $\phi$ between the N{\'{e}}el vectors of the two species (Sec.~\ref{sm:htet}), yields a fluctuational entropy, $\Delta S^{(2)}[\cos^{2}\phi]=-\frac{1}{2}\log[1-J_{ab}^{2}/J^{2}\cos^{2}\phi]\approx\frac{1}{2}J_{ab}^{2}/J^{2}\cos^{2}\phi$, favoring  collinear alignment of the N\'{e}el vectors. This indicates the observed first-order transition is driven by an order-by-disorder mechanism~\cite{Villian1980,Henley1987,Henley1989,Moessner1998} at the tetrahedral level. A more accurate estimate of the entropic advantage of the nematic state is obtained via a similar quadratic calculation for a magnetic charge-ice ground state, in which the free energy due to transverse fluctuations of AFM correlated loops with aligned N\'{e}el vectors is compared to that of loops with randomly oriented N\'{e}el vectors. The entropy of nematic order with a finite $Q_z = \langle \cos^{2}\phi\rangle-1/3$ exceeds that of randomly aligned configurations by $\frac{1}{2}Q_z J_{ab}^{2}/J^{2}$ per spin (Sec.~\ref{sm:hfull}), which is a number comparable to the entropy gain of a single tetrahedron.

Using this entropic interaction, we first investigate the possibility of loop alignment at temperatures $T$ sufficiently low such that the correlation length $L_T = J/T$ of a Heisenberg chain exceeds the length of a loop $i$. In this regime, it is well characterised by its N\'{e}el vector $\mathbf{n}_{i}$. Two such loops, $i$ and $j$, therefore experience the entropic interaction $-TN_{ij}\Delta S^{(2)}[(\mathbf{n}_{i}\cdot\mathbf{n}_{j})^{2}]$, where $N_{ij}$ is the number of shared tetrahedra, and whose leading order term is $-TN_{ij}\frac{1}{2}(J_{ab}/J)^2(\mathbf{n}_{i}\cdot\mathbf{n}_{j})^{2}$. Here $(J_{ab}/J)^2$ can be viewed as the entropic coupling parameter. For the simple loop connectivity of the ordered charge-ice structure, a mean-field description may be developed (Sec.~\ref{sm:mft1}) in which each equivalent loop is embedded in a symmetry breaking quadrupolar field, $Q_{z}=\langle n_{z}^{2}\rangle-1/3$. Requiring self-consistency then gives the temperature independent condition on the loop length, $l>l_{0}\approx13.5 (J/J_{ab})^{2}$ for $Q_{z}$ to become non-zero. For longer loops the entropy gain $l  \Delta S^{(2)} \Delta Q_z$  ($\Delta Q_z \approx 0.2$) from partial alignment overcompensates the entropy lost, $O(1 k_B)$, from constraining the fluctuations of the loop N\'{e}el vectors with the nematic order emerging discontinuously at loop lengths $l=l_{0}$. Thus, the nematic phase disappears in charge-ice structures with too short loops and/or too weak entropic couplings. This also suggests that in an ordered structure of alternating short loops and winding loops of the size of the system, the nematic transition is strongly suppressed due to the strong fluctuations of the small loops. This is indeed the case for the ordered charge-ice system II (see methods and Fig.~\ref{figOrderedStructures}b) where no signature of the nematic transition is seen (Fig.~\ref{figsm2}). The presence of the nematic phase and the value of $T_c$ are thus both sensitive to the distribution of loop length and their intertwining (connectivity). The fact that short loops tend to fluctuate strongly also rationalizes the ordering tendency in a general charge-ice structure, in particular the results of Fig.~\ref{fig3}, where progressively smaller loops align as $J/|J_{ab}|$ increases towards -1. 

The density of normal mode frequencies of the linearized charge-ice Hamiltonian shows the 1D spin-chain asymptotic form $\rho(\varepsilon)~\sim1/\sqrt{\varepsilon}$ for small $\varepsilon$ (Fig.~\ref{figsm4}), motivating a 1D Heisenberg spin chain Hamiltonian in the presence of a symmetry breaking mean-field anisotropy term, $-Q_{z}J_{ab}^{2}/J^{2}(s_{zi}^{2}-1/3)$ (see Sec.~\ref{sm:mft2}). In the long-wavelength continuum limit, this is solved exactly via numerical transfer-integral methods~\cite{McGurn1975}, predicting a first order nematic transition at $T_{0}/J\approx0.05 c J_{ab}^{2}/J^2$ where $c\lesssim 1$ represents the overall fraction of tetrahedra touched by two loops long enough to undergo nematic alignment. Note that at $T_0$, the correlation length $L_T(T_0)$ becomes of order $l_0$, such that entropy gained from nematic alignment compensates the entropy lost by constraining the fluctuations of correlated loop segments. This rationalizes the observed temperature scale of $T_{0}$ and its decrease as $J/|J_{ab}|$ becomes more negative. Indeed, for $J/|J_{ab}|=-2$ the mean-field prediction gives $T_{0}=0.0125c$, which is remarkably close to that of simulation (Fig.~\ref{fig2}) for $c\approx0.95$ which, for charge-ice, is the approximate fraction of sites involved in loops longer than $l_{0}$. Mean-field theory also explains the anomalous temperature dependence of the heat capacity in the nematic phase, giving $c_V\approx 1  -3/4\sqrt{TJ_{ab}^2/J^3}$ (Figs.~\ref{fig1}d and \ref{fig3}a) and tracing it to the quenching of the entropy of the softest spin waves due to the increasingly strong entropic interaction (Sec.~\ref{sm:mft3}). 

An $s_{z}^{2}$ anisotropy does not break time-reversal symmetry and no long range spin order is expected within the nematic phase. However, such a spontaneously emerging anisotropy reduces the $O(3)$ global symmetry of the Heisenberg Hamiltonian to an Ising $Z_{2}$ symmetry, which remains unbroken on the chains in accord with the Mermin-Wagner theorem~\cite{Mermin1966}. Above $T_0$, loops fluctuate and equilibrate rapidly due to long wavelength spin waves, whereas below $T_0$ the reduced spin symmetry entails a many orders of magnitude larger spin relaxation time due to the tiny Gibbs factor  $\exp(-J/T)$ associated with the nucleation and separation of an Ising domain-wall. With such kinetics nearly frozen out, the loops maintain their nearly perfect AFM order for very long times, with spin relaxation times of order $\tau_1 \sim \exp(J/T)$.

{\bf Conclusions and outlook.} While discontinuous transitions were found in related frustrated systems upon perturbing homogeneously the interactions and thereby lifting the ground state degeneracy~\cite{Pickles2008,Chern2008,Conlon2010,Hizi2009}, those are driven by the essentially local competition between energy and entropy. In contrast, charge-ice establishes a complex connectivity among strongly correlated non-local cluster degrees of freedom, which reflects the precise realisation of the correlated disorder --- and it is with respect to these degrees of freedom that the first order transition takes place. The predicted spin nematic breaks spin rotation symmetry, but preserves (statistical) lattice symmetries. It is thus quite distinct from lattice nematics, that break lattice rotation invariance at the level of the spin-spin correlation function~\cite{Samarakoon2022,Hallen2023}. The continuous rotational symmetry of the Heisenberg Hamiltonian is reduced to a discrete Ising symmetry, entailing an emergent slow dynamics and a new type of sudden spin-liquid freezing, in which sufficiently large loops fall out of equilibrium and become AFM ordered on mesoscopic timescales. This differs strongly from  the effect of random couplings, which may induce glassy spin freezing~\cite{Saunders2007}, with slow dynamics deriving from a complex energy landscape, but occurring at temperatures far below the dominant exchange energy scale.

Our work shows that correlated structural disorder can produce non-trivial behavior due to the emergence of non-local degrees of freedom tied to lower-dimensional clusters (loops/strings).  Solids in which similarly correlated disorder is known (or expected) to exist are numerous~\cite{Keen2015}, with corner-sharing tetrahedra being only one example of a more general class of materials whose corner or edge-sharing plaquettes may show qualitatively different magnetic behavior~\cite{Henley1989,Moessner2001,Calder2013}. Moreover, transferring the paradigm of interacting non-local intertwined magnetic degrees of freedom that arise from correlated disorder to the realm of continuous phase transitions might offer the possibility of entirely new universality classes~\cite{Halperin1983}.

Quantitatively understanding the relation between such correlated structural disorder and emergent collective degrees of freedom and their thermodynamic signatures is a formidable but not intractable problem. Indeed, experimentally observing the predicted nematic phase transition through magnetic birefringence would give indirect evidence for the existence of large loops and the presence of correlated disorder. Moreover, if it is possible to vary the exchange constants, either chemically or through a global distortion, and monitor the transition temperature and the order parameter magnitude, one might extract additional information on the distribution of loop lengths, establishing an experimental link between correlated disorder and the thermodynamics it entails.

{\bf Acknowledgments} The authors wish to thank Sam Garratt, Afonso Dos Santos Rufino, and Hugo Bocquet for helpful discussions. We also thank Christian R\"uegg for doctoral supervision of AH. The work is supported by the European Union Horizon 2020 research and innovation program under the Marie Skodowska-Curie Grant agreement No. 884104 (PSI-FELLOW-III-3i) and the Swiss National Science Foundation (grant number 200020\_182536).

{\bf Author Contributions} PMD and TF instigated the project; AH, PMD, KE and MT performed the simulations; AH, KE, TF, and PMD carried out the analyses; PMD and MM made the theoretical calculations; PMD, TF and MM wrote the paper with input from the other authors.

{\bf Methods: Monte Carlo} A single-site Monte Carlo approach was found to be sufficient for the present work. Since a wide range of temperature scales are probed, the Monte Carlo heat bath algorithm was found to be most suitable. Here, an MC move entails randomly selecting a site and calculating exactly the probability density function for that spin with all other spins fixed. This distribution is then sampled to find the new state of the chosen spin. Whilst there is a computational cost in sampling this distribution, it has the advantage of all moves being accepted and of automatically reducing the scale of variations in spin as the temperature is decreased. For more details see, for example, Ref.~\cite{Miyatake1986}.

{\bf Methods: Sample Creation} To produce a pyrochlore sample satisfying the charge-ice constraint on each tetrahedron, the pyrochlore lattice of size $L$ (containing $16L^{3}$ atoms) is constructed and initially populated with $a$ and $b$ sites according to an ordered structure consisting of $[110]$ and $[1\bar{1}0]$ chains of sites of one or the other type of cation, respectively. Under periodic boundary conditions, this may be seen as a regular array of $4L^{2}$ winding loops of length $4L$. The connectivity of such a structure is characterized by any two loops sharing either zero or one tetrahedron. This initial structure will be referred to as an ordered charge-ice system I. To disorder it, a loop consisting of alternating site types is identified via a worm algorithm and all site types are interchanged, preserving the charge-ice structure. This procedure is repeated until variations in loop structure satisfy the known statistical properties of the loops as detailed in Figs.~\ref{fig1}b and in Ref.~\cite{Jaubert2011}. These samples will be referred to as a charge-ice system. An alternative ordered charge-ice system may be constructed consisting of $(100)$ planes of $[110]$ chains of sites separated by regions fully populated by hexagonal loops of length $l=6$. This is referred to as the ordered charge-ice system II and contains $L^2$ loops of length $4L$ and $2L^3$ of length 6. See Fig.~\ref{figOrderedStructures} which visualizes both charge ordered systems.

\end{bibunit}

\renewcommand{\thefigure}{SM\arabic{figure}}
\setcounter{figure}{0}

\renewcommand{\theequation}{SM\arabic{equation}}
\setcounter{equation}{0}

\renewcommand{\thesection}{SM\arabic{section}}
\setcounter{section}{0}

\begin{bibunit}

\section{Supplementary Material}

\subsection{Breaking up the giant loops}

\begin{figure*}
	\includegraphics[width=0.8\linewidth,trim=0cm 0cm 0cm 0cm,clip]{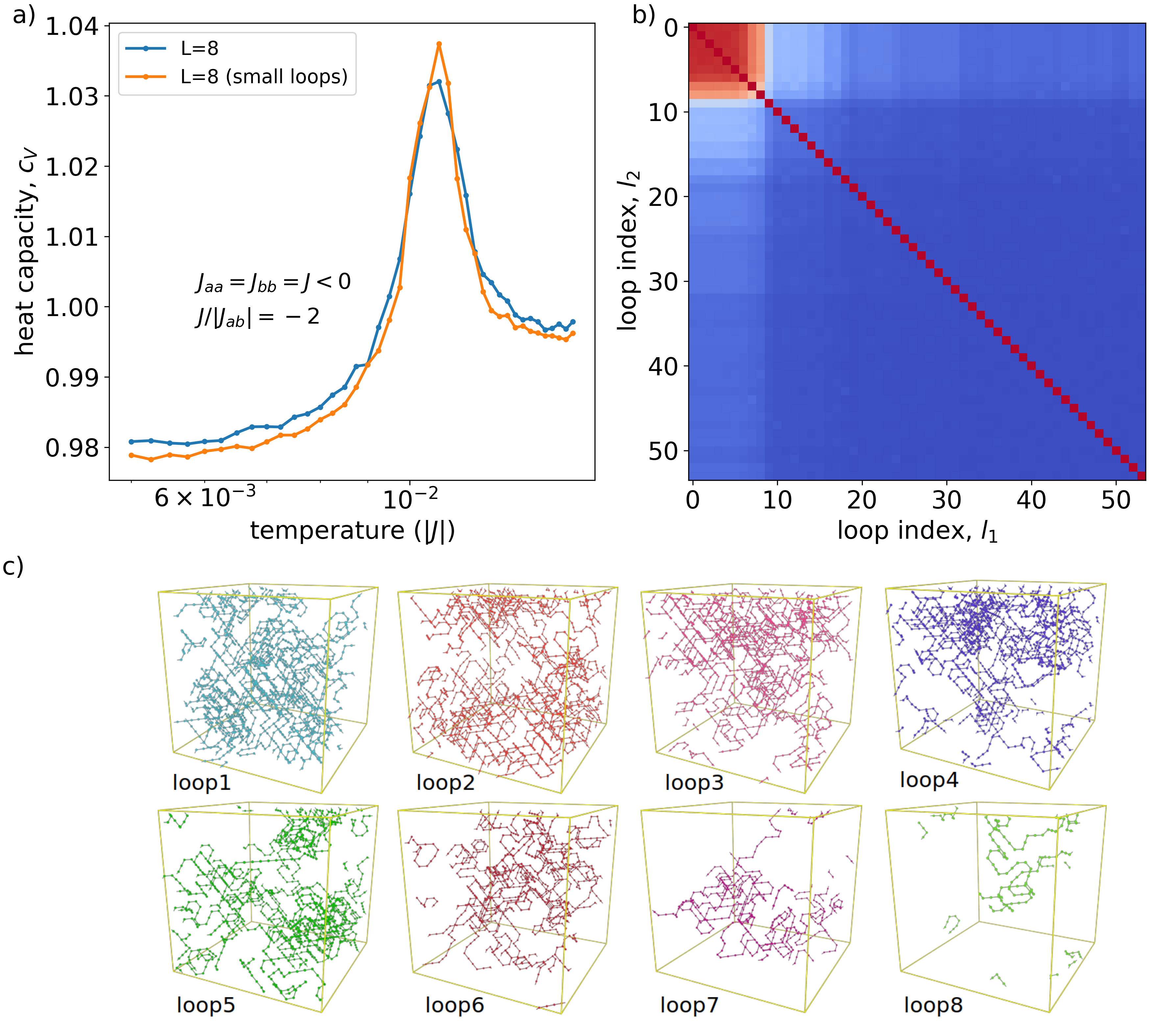}
	\caption{{\bf Breaking up the giant loops.} a) Heat capacity and b) loop-loop quadrupolar correlation of an $L=8$ charge-ice system with $J/|J_{ab}|=-2$ in which the giant loops are decomposed into smaller winding loops. c) Visualization of its eight largest loops.}
	\label{figsm1}
\end{figure*}

To investigate the robustness of the observed phase transition with respect to the size of the giant loops, we perform Monte Carlo simulations on an $L=8$ system for which the loop structure generation was biased to generating smaller loops. This bias was achieved by only allowing changes in the structure which reduced the sum of the square of loop lengths. In particular, this procedure was applied to the $L=8$ sample used in the main text, resulting in a sample (referred to as the ``small loop'' sample) with over 54 loops, the largest ten of which had lengths 1386, 1376, 1236, 1218, 1058, 826, 426, 140, 86, and 30. This should be compared with the original sample which had 35 loops, the largest four of which are 3926, 3908, 60 and 26 in length. Fig.~\ref{figsm1}a displays the resulting heat capacity compared to the original $L=8$ charge-ice system showing little change in the transition temperature $T_{0}$. Fig.~\ref{figsm1}b displays the loop-loop orientation correlation $\langle{\rm Tr}(\mathbf{\hat{Q}}_{l_1}	\mathbf{\hat{Q}}_{l_2})\rangle/(\langle|\mathbf{Q}_{l_{1}}|^{2}\rangle\langle|\mathbf{Q}_{l_{2}}|^{2}\rangle)^{1/2}$ below the critical temperature, demonstrating that the growth in the bulk quadrupolar order parameter is due to the alignment of these larger non-giant loops. Fig.~\ref{figsm1}c visualizes the 8 largest loops of the sample, showing that all but the eighth largest loop are winding.

\subsection{Ordered charge-ice}

\begin{figure*}
	\includegraphics[width=0.8\linewidth,trim=1.5cm 4cm 2.5cm 4cm,clip]{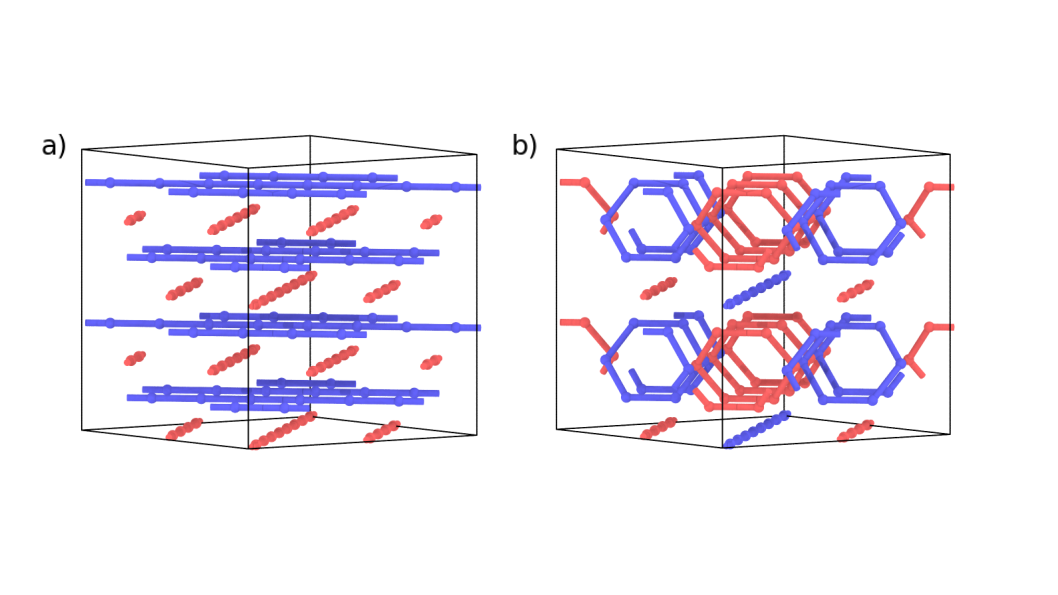}
	\caption{{\bf Ordered charge ice structures.} The two ordered charge ice structures, a) system I and b) system II. The loops can be identified via similarly coloured bonds with the colour also reflecting the cation type. In system I, each linear loop will share either zero or at most one tetrahedron with any other linear loop. In system II, the linear loops do not share tetrahedra between themselves.}
	\label{figOrderedStructures}
\end{figure*}	

\begin{figure*}
	\includegraphics[width=0.8\linewidth,trim=0cm 0cm 0cm 0cm,clip]{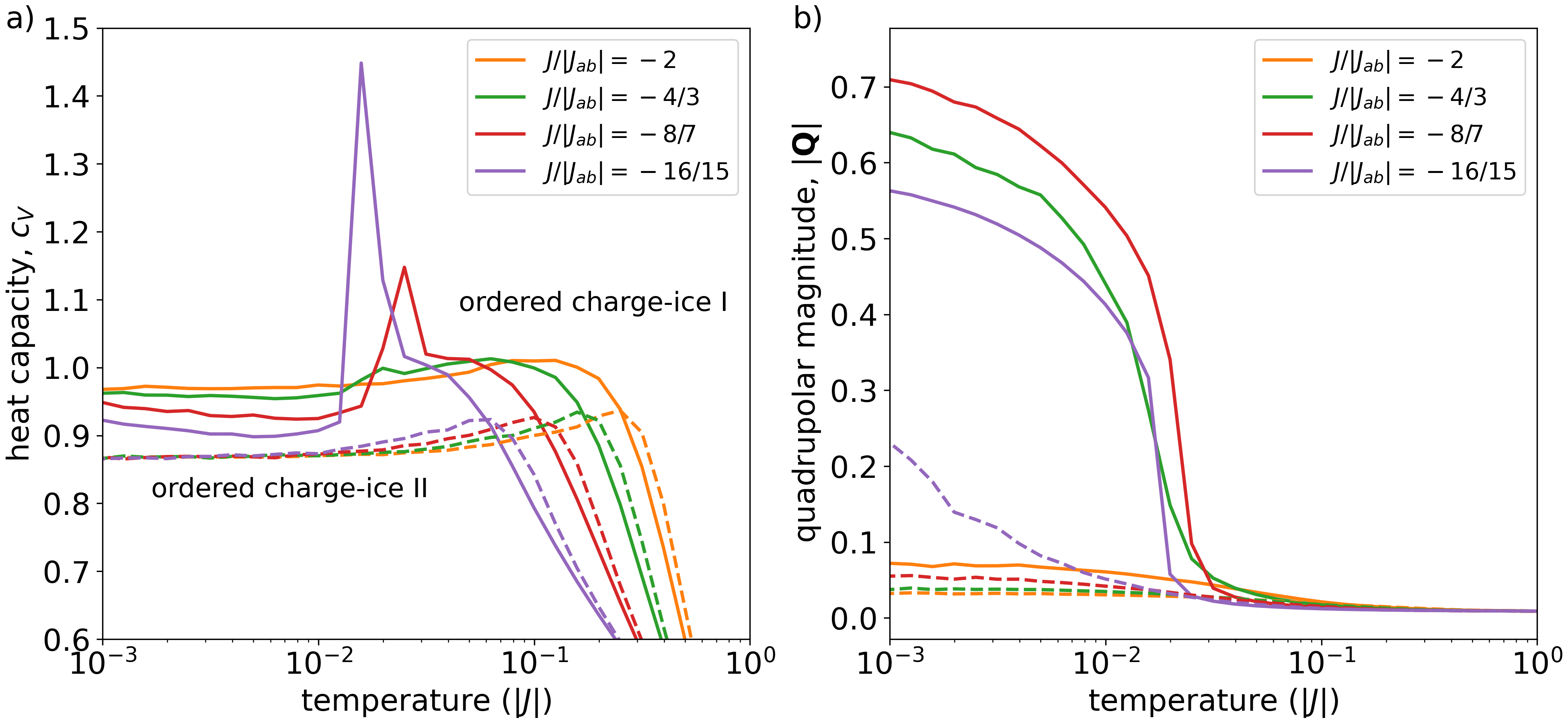}
	\caption{{\bf Thermodynamics of ordered charge ice structures.} a) Heat capacity and b) average quadrupolar magnitude of the ordered charge-ice system I (solid lines) and of the ordered charge-ice system II (dashed lines) for $L=8$ and a range of $J/|J_{ab}|$.}
	\label{figsm2}
\end{figure*}

Fig.~\ref{figOrderedStructures} displays the two ordered charge ice systems I and II for the case of $L=8$. Both structures are investigated for inter-chain couplings $J/|J_{ab}|$ equal to -2, -4/3, -8/7, -16/15. Fig.~\ref{figsm2} displays a) the heat capacity and b) the average magnitude of the bulk quadrupolar order parameter for system I showing that for $J/|J_{ab}|=-2$ the first order phase transition is entirely suppressed, as predicted by mean field theory (Sec.~\ref{sm:mft1}). However as $J/|J_{ab}|$ becomes less negative, the transition appears, with $T_{0}$ again increasing as $J/|J_{ab}|$ approaches -1, as for the case of the general charge-ice structure (Figs.~1-3 in the main text). Fig.~\ref{figsm2} also shows similar data for the ordered charge-ice system II where due to the linear loops not sharing any tetrahedra, interacting only via hexagonal loops of length six that do not align, the nematic transition is generally absent for all choices of $J/|J_{ab}|$ except very close to the PHAFM case where the hexagons also begin to align. For both ordered charge-ice structures, when no transition is observed, the heat capacity plateaus to a value equal to $1-N_{l}/(16L^2)$, where $N_{l}$ is the number of loops. This originates from the zero modes within the system~\cite{Moessner1998}, which for charge ice is equal to twice the number of loops. For ordered charge ice I, $N_{l}=4L^2$ and for ordered charge ice II $N_{l}=L^2+2L^3$ giving the respective heat capacity plateaus of 0.97 and 0.87.

\subsection{Harmonic transverse spin fluctuations} \label{sm:harmonic}

The classical Heisenberg spin Hamiltonian may be written as
\begin{equation}
	H=-\frac{1}{2}\sum_{ij}J_{ij}\hat{\mathbf{s}}_{i}\cdot\hat{\mathbf{s}}_{j}, \label{EqnSM1}
\end{equation}
for which the local field at each site $i$ is 
\begin{equation}
	\mathbf{b}_{i}=\sum_{i}J_{ij}\hat{\mathbf{s}}_{j}.
	\label{EqnSM2}
\end{equation}

To investigate the transverse spin fluctuations, $\mathbf{s}_{\perp,i}$ with respect to a given spin configuration $\hat{\mathbf{s}}_{0,i}$ each spin is written as
\begin{equation}
	\hat{\mathbf{s}}_{i}=\mathbf{s}_{\perp,i}+\hat{\mathbf{s}}_{0,i}
	|{\mathbf{s}}_{\|,i}|
	\label{EqnSM3}
\end{equation}
where $|\mathbf{s}_{\|,i}|=\sqrt{1-|\mathbf{s}_{\perp,i}|^{2}}$. If the magnetic configuration $\hat{\mathbf{s}}_{0,i}$ is at a local energy minimum then all $\hat{\mathbf{s}}_{0,i}$ will be parallel to their local fields, $\mathbf{b}_{0,i}$. Then to quadratic order in the transverse components, the Hamiltonian may be written as $H=H^{(0)}+\Delta H^{(2)}$ where
\begin{eqnarray}
	\Delta H^{(2)}&=&-\frac{1}{2}\sum_{ij}\left(J_{ij}-|\mathbf{b}_{0,i}|\delta_{i,j}\right)	\mathbf{s}_{\perp,i}\cdot\mathbf{s}_{\perp,j}\nonumber\\
	&=&-\frac{1}{2}\sum_{ij}\Lambda_{ij}	\mathbf{s}_{\perp,i}\cdot\mathbf{s}_{\perp,j}.
	\label{EqnSM4}
\end{eqnarray}
In the above, the off-diagonal term $J_{ij}$ is the full 3D Hessian whereas the second diagonal term is a correction to the 3D Hessian which projects the taken derivatives onto the tangent space of each spin. 

Representing the 2D tangent space of spin $i$ as $\mathbf{e}_{1,i}$ and  $\mathbf{e}_{2,i}$ with $\mathbf{e}_{1,i}\times\mathbf{e}_{2,i}=\hat{\mathbf{s}}_{0,i}$, the $i$th spin may be written as
\begin{equation}
	\mathbf{s}_{\perp,i}
	=\sum_{\alpha=1,2}\chi^{\alpha,i}\mathbf{e}_{\alpha,i}
	\label{EqnSM5}
\end{equation}
where $\chi^{\alpha,i}$ are real numbers. The choice of $\mathbf{e}_{\alpha,i}$ is not unique and we follow Ref.~\cite{Mueller2018}. Together the above yields a symmetric matrix $M_{i\alpha,j\beta}$ of rank $2N$, represented as an $N\times N$ matrix of $2\times2$ block elements, whose $(i,j)$th block element is $\Lambda_{ij}\mathbf{e}_{\alpha,i}\cdot\mathbf{e}_{\beta,j}$.  Solving the corresponding eigen-problem yields the normal modes of Eqn.~\ref{EqnSM4} that govern the fluctuations of this quadratic Hamiltonian. It is noted that for disordered/frustrated systems, $\hat{\mathbf{s}}_{0,i}$ and thus the local tangent space, defined via $\mathbf{e}_{\alpha,i}$, will be different for each spin. Thus the normal modes presently calculated are non-trivially related to the corresponding spin-wave modes which arise from a linearisation of the Landau-Lifschitz equation.

At the level of the quadratic approximation to the Hamiltonian, the resulting partition function becomes a simple Gaussian integral, evaluating to
\begin{equation}
	Z=\prod_{n|\lambda_n >0}\sqrt{\frac{2\pi T}{\lambda_{n}}},
\end{equation} 
where the $\lambda_{n}$ are the non-zero eigenvalues of the fluctuation matrix $M$, from which the free energy may be calculated as $F=-T\log Z$ giving
\begin{equation}
	F=\frac{T}{2}\sum_{n| \lambda_n >0}\log\frac{\lambda_{n}}{2\pi T}.
\end{equation}
In the thermodynamic limit this can be evaluated as an integral $\int_{0^+}^{\infty}\mathrm{d}\lambda\,\rho(\lambda)\log\lambda$ using the density of eigenvalues (or density of states DOS), $\rho(\lambda)$, which is normalized to $2N$.

\subsubsection{Single tetrahedron} \label{sm:htet}

The Hamiltonian for a single tetrahedron satisfying the charge-ice rule is given by
\begin{eqnarray}
	H&=&-J_{aa}\mathbf{\hat{s}}_{a,1}\cdot\mathbf{\hat{s}}_{a,2}-J_{bb}\mathbf{\hat{s}}_{b,1}\cdot\mathbf{\hat{s}}_{b,2}+\nonumber\\
	& &-J_{ab}\left(\left(\mathbf{\hat{s}}_{a,1}+\mathbf{\hat{s}}_{a,2}\right)\cdot
	\left(\mathbf{\hat{s}}_{b,1}+\mathbf{\hat{s}}_{b,2}\right)\right),
\end{eqnarray}
where we recall that we focus on the parameter regime where the $J_{a}$ and $J_{b}$ couplings are negative (AFM). For a ground state configuration of region IV, we have the AFM configurations between spins of the same type: $\mathbf{\hat{s}}_{a,1}=-\mathbf{\hat{s}}_{a,2}$ and $\mathbf{\hat{s}}_{b,1}=-\mathbf{\hat{s}}_{b,2}$, and an angle $\phi$ between the alignment axis. This gives the ground state energy $-J_{aa}-J_{bb}$ independent of $\phi$. Using the formalism of the previous section, the quadratic Hamiltonian is represented as a matrix of rank 8:
\begin{widetext}
	\begin{equation}
		M_{i\alpha,j\beta}=
		\begin{bmatrix}
			\left| J_{aa}\right|  & 0 & -J_{aa} & 0 & -J_{ab} & 0 & -J_{ab} & 0 \\
			0 & \left| J_{aa}\right|  & 0 & J_{aa} & 0 & -J_{ab}\cos\phi & 0 & J_{ab}\cos\phi \\
			-J_{aa} & 0 & \left| J_{aa}\right|  & 0 & -J_{ab} & 0 & -J_{ab} & 0 \\
			0 & J_{aa} & 0 & \left| J_{aa}\right|  & 0 & J_{ab}\cos\phi & 0 & -J_{ab}\cos\phi \\
			-J_{ab} & 0 & -J_{ab} & 0 & \left| J_{bb}\right|  & 0 & -J_{bb} & 0 \\
			0 & -J_{ab}\cos\phi & 0 & J_{ab}\cos\phi & 0 & \left| J_{bb}\right|  & 0 & J_{bb} \\
			-J_{ab} & 0 & -J_{ab} & 0 & -J_{bb} & 0 & \left| J_{bb}\right|  & 0 \\
			0 & J_{ab}\cos\phi & 0 & -J_{ab}\cos\phi & 0 & J_{bb} & 0 & \left| J_{bb}\right|  \\
		\end{bmatrix},
	\end{equation}
\end{widetext}
whose four non-zero eigenvalues give the free energy contribution
\begin{equation}
	\Delta F^{(2)}[T,\phi]= \frac{T}{2}\log\left[\frac{\left(J_{\mathrm{aa}}J_{\mathrm{bb}}-J_{\mathrm{ab}}^{2}\right)\left(J_{\mathrm{aa}}J_{\mathrm{bb}}-J_{\mathrm{ab}}^{2}\cos^{2}\phi\right)}{\pi^{4}T^{4}}\right]. \nonumber
\end{equation}
The above can be conveniently written as
\begin{equation}
	\Delta F^{(2)}[T,\phi]=\Delta F^{(2)}[T]-T\Delta S^{(2)}[\cos^{2}\phi]
\end{equation}
where (with $k_{\mathrm{B}}=1$) 
\begin{equation}
	\Delta S^{(2)}[\cos^{2}\phi]=-\frac{1}{2}\log\left[1-\frac{J_{ab}^{2}\cos^{2}\phi}{J_{aa}J_{bb}}\right] \label{EqnSM6a}
\end{equation}
is the (temperature independent) fluctuational entropy evaluated for a given angle $\phi$ between the orientations of the two equal species  pairs. Thus the angle-constrained free energy has minima at $\phi=0,\pi$ and maxima at $\phi=\pm\pi/2$. Alignment or anti-alignment thus results in maximal fluctuational entropy, where $\Delta S^{(2)}\equiv \Delta S^{(2)}(\phi=0)-\langle\Delta S^{(2)}(\phi)\rangle_\phi\approx 0.1$ for $J_{ab}/\sqrt{J_{aa}J_{bb}}=1/2$.

\subsubsection{Full system} \label{sm:hfull}

A similar harmonic analysis may be carried out for the full charge-ice system, where now the eigen-system of the Hessian $M$ (calculated via Eqn.~\ref{EqnSM4}) must be solved numerically for a particular choice of the $T=0$ reference configuration.

\begin{figure}
	\includegraphics[width=0.7\linewidth,trim=0cm 0cm 0cm 0cm,clip]{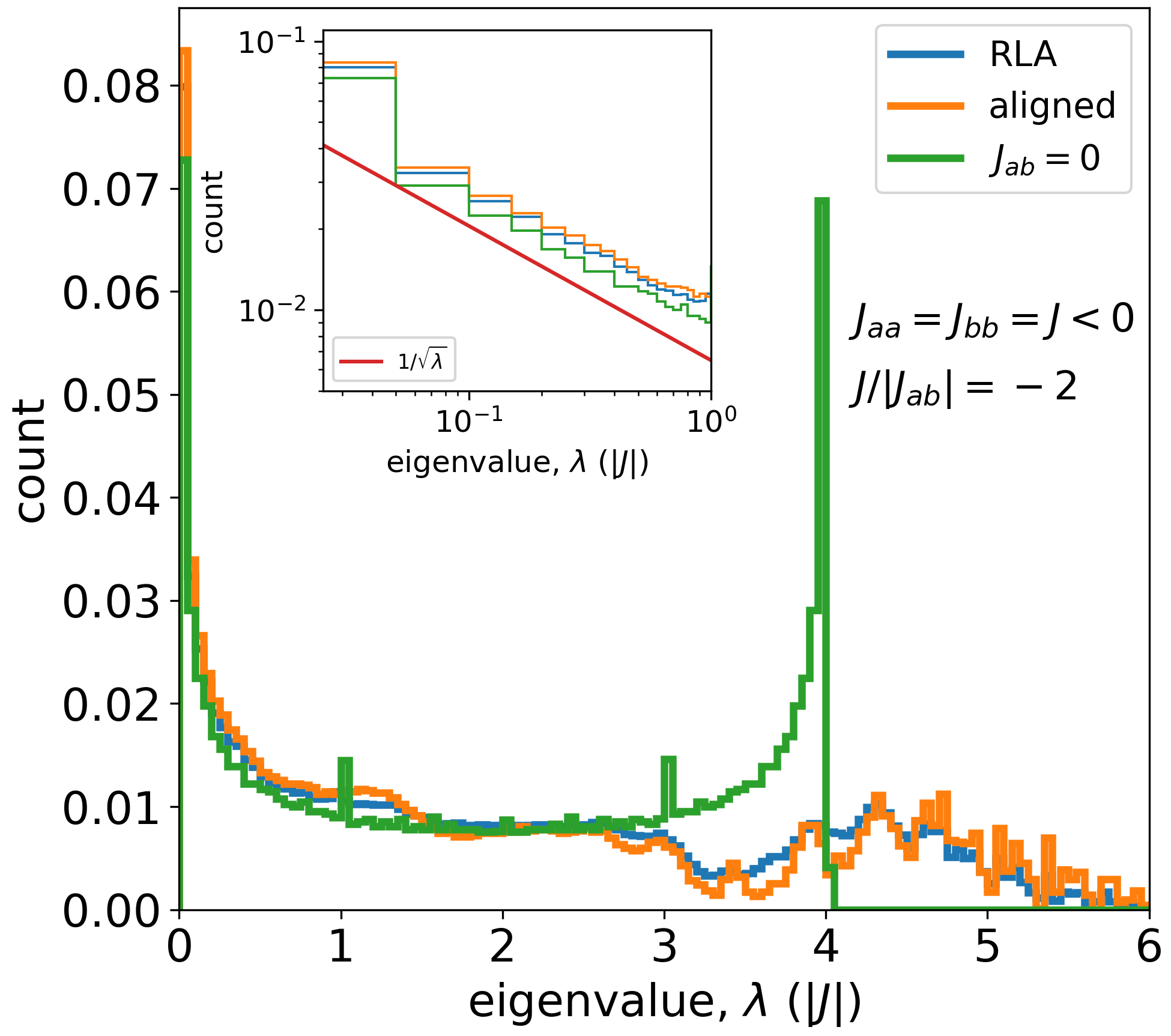}
	\caption{{\bf Density of normal mode frequencies of the transverse harmonic spin Hamiltonian.} Data is shown for both  aligned and randomly aligned (RLA) ground state configurations, respectively. Each loop contributes two zero-modes (not shown). In the low frequency regime, a power-law $1/\sqrt{\lambda}$ is seen, suggesting the dominance of 1D long-range-ordered AFM spin-chain behaviour.} \label{figsm4}
\end{figure}

Fig.~\ref{figsm4} displays the normal mode density of states (DOS) for two ground-state configurations: one with nematic order, in which all loop N\'{e}el vectors are aligned; and one where they are randomly orientated with respect to each other (random loop AFM or RLA).  These states are both members of the manifold of ground states identified by Banks and Bramwell~\cite{Banks2012}, and are indistinguishable in terms of their internal energies. The DOS of the RLA depends somewhat on the particular ground state configuration, but for sufficiently large samples self-averaging reduces such differences. For smaller samples an average over many choices of random alignment results in a converged DOS. Both RLA and nematic order reveal $2N_{\mathrm{loop}}$ zero-modes reflecting the individual $O(3)$ symmetry of each AFM loop, and whilst there are differences between the nematic and random ground state configurations (e.g. the enhanced density of low frequency states and more discrete structure at higher frequencies in the nematic ground state), the similarities at low frequency are more revealing. In particular, a log-log plot (inset of Fig.~\ref{figsm4}) reveals the asymptotic form $\rho(\lambda)~\sim1/\sqrt{\lambda}$ for small $\lambda$ -- a hallmark signature of the fluctuation spectrum of  AFM-ordered reference configurations of 1D  spin chains. For comparison the DOS derived from the harmonic Hessian with $J_{\mathrm{ab}}=0$ is also shown, which consists of $N_{\mathrm{loop}}$ non-interacting finite 1D antiferromagnetic spin chains. This is in agreement with the known analytical form (apart from finite size corrections due to a small fraction of short loops) $\rho(\lambda)\propto \sqrt{\lambda(4J-\lambda)}$. 

Within the harmonic approximation, the difference in fluctuational entropy between nematic and RLA states is given by  $-\int_{0^+}^{\infty}\mathrm{d}\lambda\,\left(\rho_{\mathrm{nematic}}(\lambda)-\rho_{\mathrm{RLA}}(\lambda)\right)\ln(\lambda)$ which we find to be a positive quantity. This originates from the coupling-induced softening of low frequency normal modes, which enhances the $1/\sqrt{\lambda}$ tail. This softening effect is strongest for the nematically aligned configuration (hence the enhancement at low frequency regime compared to a randomly aligned ground state configuration), which is thus entropically favored. 

\subsection{Meanfield theory of a nematic transition}

\subsubsection{Nematic alignment} \label{sm:mft1}

In what follows, we consider loops of characteristic length $l$ at low temperatures $T\ll|J|/l$, such that the persistence (or correlation) length of an infinite Heisenberg chain $L_{T}=|J|/T\gg l$. In this limit, each loop $i$ can be characterised by a single N\'{e}el vector, $\mathbf{\hat{n}}_{i}$ (which may still flip slowly), and exhibits fast but small transverse fluctuations around it. From Sec.~\ref{sm:htet} the total free energy of an ensemble of such loops is
\begin{equation}
	-\frac{1}{2}\sum_{i,j}N_{ij}T\Delta S^{(2)}\left[(\mathbf{\hat{n}}_{i}\cdot\mathbf{\hat{n}}_{j})^{2}\right], \label{eqnsm}
\end{equation}
where $N_{ij}$ is the number of shared tetrahedra between loops $i$ and $j$. This gives the temperature independent partition function,
\begin{equation}
	Z\left[\{\mathbf{n}_{i}\}\right]=\exp\left[\frac{1}{2}\sum_{i,j}N_{ij}\Delta S^{(2)}\left[(\mathbf{\hat{n}}_{i}\cdot\mathbf{\hat{n}}_{j})^{2}\right]\right].
\end{equation}

As a specific example we consider a periodic charge-ordered system of size $L$, which contains $N_{l}=4L^2$ loops of length $l=4L$, where each loop shares either zero or one tetrahedron with any other loop. Thus $N_{ij}=1$ and
\begin{equation}
	Z\left[\{\mathbf{n}_{i}\}\right]=\exp\left[\frac{1}{2}\sum_{<i,j>}\Delta S^{(2)}\left[(\mathbf{\hat{n}}_{i}\cdot\mathbf{\hat{n}}_{j})^{2}\right]\right].
\end{equation}
where the $j$ summation spans the $l$ loops that share a tetrahedron with the $i$th loop. A mean-field construction is performed by replacing the summand by an average with respect to $\mathbf{\hat{n}}_{j}$, yielding an effective single loop weight:
\begin{eqnarray}
	Z^{\mathrm{\rm MF}}_{i}(\mathbf{\hat{n}}_{i})&=&\exp\left[l\left\langle\Delta S^{(2)}\left[(\mathbf{\hat{n}}_{i}\cdot\mathbf{\hat{n}}_{j})^{2}\right]\right\rangle_{\mathbf{\hat{n}}_{j}}\right]\nonumber\\
	&\approx&\exp\left[\frac{lJ_{ab}^{2}}{2J^{2}}\left\langle(\mathbf{\hat{n}}_{i}\cdot\mathbf{\hat{n}}_{j})^{2}\right\rangle_{\mathbf{\hat{n}}_{j}}\right],
\end{eqnarray}
where in the last equality we have expanded $\Delta S^{(2)}\left[(\mathbf{\hat{n}}_{i}\cdot\mathbf{\hat{n}}_{j})^{2}\right]$ to leading order in $J_{ab}$, assuming $J_{aa}=J_{bb}=J$.

We now determine whether it is consistent to assume that dyads of $\mathbf{\hat{n}}_{i}$ acquire a finite expectation value. Choosing the $z$ axis as the symmetry breaking axis we assume $Q_{z}=\langle n_{z}^{2}\rangle-1/3$, where a non-zero value would spontaneously break the rotational invariance. Assuming rotational symmetry around the $z$-axis, $\langle n_{x}^{2}\rangle=\langle n_{y}^{2}\rangle=(1-\langle n_{z}^{2}\rangle)/2$, and $\langle n_{x}\rangle=\langle n_{y}\rangle=0$, the above average evaluates to 
\begin{equation}
	\left\langle(\mathbf{\hat{n}}_{i}\cdot\mathbf{\hat{n}}_{j})^{2}\right\rangle_{\mathbf{\hat{n}}_{j}}=\frac{3}{2}Q_{z}n_{zi}^{2}.
\end{equation}
Self consistency of the mean field now requires that $\langle n_{z}^{2}\rangle-1/3$ computed with the effective single loop weight 
\begin{equation}
	Z^{\mathrm{MF}}_{i}=\exp\left[\frac{3}{4}l(J_{ab}/J)^{2}Q_{z}n_{z,i}^{2}\right]
\end{equation}
equal $Q_{z}$. Calling $\lambda=3(J_{ab}/J)^{2}/4$, we thus seek the stable solution of the mean field equation:
\begin{equation}
	Q_{z}=\frac{\int_{-1}^{1}\mathrm{d}n_{z}\, \exp[\lambda l Q_{z} n^2_{z}](n_{z}^{2}-1/3)}{\int_{-1}^{1}\mathrm{d}n_{z}\, \exp[\lambda l Q_{z} n^2_{z}]}. \label{eqnsm1}
\end{equation}
For large $\lambda$, $Q_{z}$ tends to $2/3$. This symmetry breaking solution disappears at $l \lambda\approx10.1=(l\lambda)_{0}$ where the order parameter discontinuously drops from $Q_{z}\approx0.205$ at $(l\lambda)_{0}$ to zero, signaling a first order transition. It is noted that a spinodal instability of the disordered phase exists at $(l\lambda)_{sp}=45/4=11.25$, which is however preempted by the above first order transition --- as required for a nematic transition, which cannot be continuous. 

Low temperature order is thus predicted to exist only for loops larger than $l_{0}= (l\lambda)_{0}/\lambda= 4(l\lambda)_{0}(J/J_{ab})^{2}/3$. For the case of our ordered charge-ice where $l=4L$, the nematic phase transition will only occur for periodic samples of size $L$, when  $J_{ab}/|J|>J_{ab}^{0}/|J|=(4(l\lambda)_{0}/3l)^{1/2}=((l\lambda)_{0}/3L)^{1/2}$. For the case of $L=8$ this requires $J_{ab}/J>0.65$ or $J/|J_{ab}|<-1.53$. 

\begin{figure*}
	\includegraphics[width=0.7\linewidth,trim=0cm 0cm 0cm 0cm,clip]{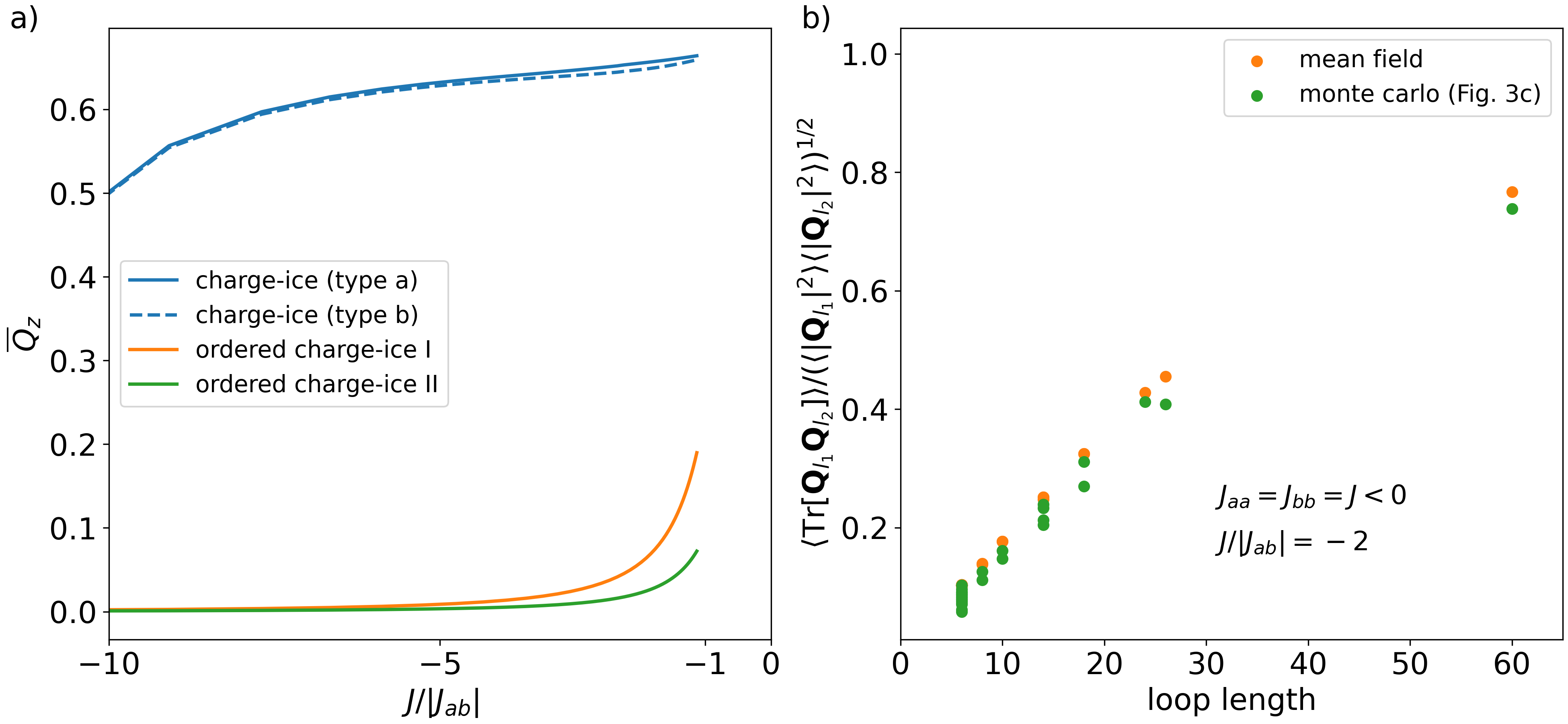}
	\caption{{\bf Mean field description of nematic alignment.} a) Self-consistent mean-field quadrupolar component for $T\to 0$ as a function of $J/|J_{ab}|$ using the discrete loop distributions of the $L=8$ charge-ice realization of Fig.~3 in the main text and the charge ordered systems of Fig.~\ref{figsm2}. It is noted that values of $J/|J_{ab}|$ nearing -1 are outside the perturbative regime presently assumed. b) Mean-field prediction of $\langle{\rm Tr}[\mathbf{Q}_{l_1}\mathbf{Q}_{l_2}]\rangle/(\langle|\mathbf{Q}_{l_{1}}|^{2}\rangle\langle|\mathbf{Q}_{l_{2}}|^{2}\rangle)^{1/2}$ for small loops ($l_1$) embedded in the quadrupolar field of larger loops. Here we take $l_2$ to be a giant loop. This is to be compared to Monte-Carlo simulations of the $L=8$ charge-ice realization of Fig.~3 in the main text for the case of $J/|J_{ab}|=-2$. } \label{figsm5}
\end{figure*}

A fully self-consistent mean-field theory with respect to normalized loop distributions for type $a$ and $b$, $P_{a/b}(l)$, now follows by writing the average quadrupolar field component for sites of type $a/b$ as
\begin{equation} 
	\overline{Q}_{z}^{a/b}=\frac{\sum_{l}l P_{a/b}(l)Q_{x}^{a/b}(l)}{\sum_{l}l P_{a/b}(l)}
\end{equation}
and
\begin{equation} 
	Q_{z}^{a/b}(l)=\frac{\int_{-1}^{1}\mathrm{d}n_{z}\, \exp[l\lambda\overline{Q}_{z}^{b/a} n^2_{z}](n_{z}^{2}-1/3)}{\int_{-1}^{1}\mathrm{d}n_{z}\, \exp[l\lambda\overline{Q}_{z}^{b/a} n^2_{z}]}.
\end{equation}
Here $Q_{z}^{a/b}(l)$ is the quadrupolar mean-field felt by loops of length $l$ of type $a/b$. Fig.~\ref{figsm5}a displays the self-consistent values of $\overline{Q}_{z}^{a/b}$ obtained upon iteration of the above, as a function of $J/|J_{ab}|$, using the discrete loop distributions for the $L=8$ ordered charge-ice (systems I and II) and the charge-ice realization used in Fig.~3 of the main text. The charge-ice configuration $\overline{Q}_{z}^{a/b}$ rapidly saturates to a maximum value, whereas for the charge-ordered structure, $\overline{Q}_{z}^{a/b}$ indeed remains small. For the former, only a small difference is seen in the average value experienced by sites of type $a$ and $b$ reflecting the similar sizes of the two giant loops, whereas for the latter ordered structures the $\overline{Q}_{z}^{a/b}$ are identical in value. For the ordered charge-ice system II, $\overline{Q}_{z}^{a/b}$ grows most weakly reflecting the large number of hexagonal loops in the system. 

The $\overline{Q}_{z}^{a/b}$ and Eqn.~\ref{eqnsm1} may be used to calculate $\langle{\rm Tr}[\mathbf{Q}_{l_1}\mathbf{Q}_{l_2}]\rangle/(\langle|\mathbf{Q}_{l_{1}}|^{2}\rangle\langle|\mathbf{Q}_{l_{2}}|^{2}\rangle)^{1/2}$, where loop $l_1$ is a giant loop whose quadrupolar field is oriented along $\mathbf{\hat{z}}$ with magnitude $\overline{Q}_{z}^{a/b}$ and $l_2$ is a small loop of type $b/a$. Fig.~\ref{figsm5}b compares this to the data of Fig.~3c showing very good agreement, and quantitatively confirming the initial assumption entailed in Eqn.~\ref{eqnsm} and the general mean-field approach. For this system $\overline{Q}_{z}^{a}/(2/3)=0.983$ and $\overline{Q}_{z}^{b}/(2/3)=0.971$, giving the effective fraction of tetrahredra participating in the nematic alignment. These numbers are comparable to the value $c\approx0.975$ obtained when $c$ is given by the fraction of tetrahedra that touch those loops participating in the nematic alignment (see main text). 

\subsubsection{Finite temperature nematic phase transition} \label{sm:mft2}

We now consider the full  loop-resolved effective spin Hamiltonian $H=\sum_{l}H_{l}$, where for each loop $l$ we have
\begin{equation}
	H_{l}=-J\sum_{i\in l} \mathbf{\hat{s}}_{i}\cdot\mathbf{\hat{s}}_{i+1}
	-T\frac{1}{2\cdot4}\sum_{i\in l}\sum_{j\in \overline{nn}(i)}
	\Delta S^{(2)}\left[(\mathbf{\hat{s}}_{i}\cdot\mathbf{\hat{s}}_{j})^{2}\right].
	\label{EqnSM16a}
\end{equation}
Here $\overline{nn}(i)$ are the $i$th site nearest neighbours of opposite type. In the above we assume that temperatures are sufficiently low that AFM order exists at the length-scale of the tetrahedron, so that the spin directions define the local orientation of the AFM structure. The factor of 1/4 takes into account that the tetrahedron free energy as calculated in Sec.~\ref{sm:htet} involves four $J_{ab}$ bonds. Expanding the orientational entropy with respect to $J_{ab}$ results in the leading order term
\begin{equation}
	H_{l}\approx-J\sum_{i\in l} \mathbf{\hat{s}}_{i}\cdot\mathbf{\hat{s}}_{i+1}
	-T\Delta S\frac{1}{2\cdot4}\sum_{i\in l}\sum_{j\in \overline{nn}(i)}
	(\mathbf{\hat{s}}_{i}\cdot\mathbf{\hat{s}}_{j})^{2}.
	\label{EqnSM16b}
\end{equation}
where
\begin{equation}
	\Delta S=\frac{1}{2}\frac{J_{ab}^2}{J^2}.
\end{equation}
We now decouple the loops in a mean field spirit, rewriting $(\mathbf{\hat s}_i\cdot\mathbf{\hat s}_j)^2=\sum_{\alpha,\beta} s_{\alpha i} s_{\beta i} s_{\alpha j} s_{\beta j}$ as
		\begin{eqnarray}
	&&\sum_{\alpha,\beta} \left[(s_{\alpha i} s_{\beta i}-\langle s_{\alpha i} s_{\beta i}\rangle)  (s_{\alpha j} s_{\beta j}- \langle s_{\alpha j} s_{\beta j}\rangle)  \right.\nonumber \\
	&&\quad\quad+ s_{\alpha i} s_{\beta i} \langle s_{\alpha j} s_{\beta j}\rangle+ s_{\alpha j} s_{\beta j} \langle s_{\alpha i} s_{\beta i}\rangle\nonumber \\
	&&\quad\quad\left.- \langle s_{\alpha i} s_{\beta i}\rangle \langle s_{\alpha j} s_{\beta j}\rangle \right]. \label{EqnSM16c}
\end{eqnarray}
We assume quadrupolar order to set in, and choosing the polarization axis along $z$ gives
\begin{equation}
	\langle s_{\alpha i} s_{\beta i}\rangle = \delta_{\alpha\beta}\left(\frac{1}{3}+Q_z\left(\frac{3}{2}\delta_{\alpha z}-\frac{1}{2}\right)\right),
\end{equation}
(such that the trace equals 1), while rotational invariance implies $\langle s_{xi} \rangle= \langle s_{yi} \rangle = \langle s_{zi} \rangle=0$. Substitution of the above into Eqn.~\ref{EqnSM16c} with the first term dropped, finally gives,
\begin{equation}
	(\mathbf{\hat s}_i\cdot\mathbf{\hat s}_j)^2\approx\frac{3}{2}Q_z\left(s_{zi}^2 -\frac{1}{3} +s_{zj}^2-\frac{1}{3}\right)-\frac{3}{2}Q_z^2+\frac{1}{3}
\end{equation}
and the mean field loop Hamiltonian:
\begin{equation}
	H_{\ell}= -\sum_{i\in \ell}\left(J \mathbf{\hat{s}}_{i}\cdot\mathbf{\hat{s}}_{i+1}
	+ D\left(s_{zi}^2-\frac{1}{3}\right)-\frac{1}{2}A T Q_z^2 \right)\label{anisotropicHeisenberg}
\end{equation}
with
\begin{equation}
	D=  \frac{3}{2}Q_z T\Delta S \equiv A T Q_z , \quad A\equiv \frac{3}{2}\Delta S.
\end{equation}

When performing the simple gauge transformation from AFM to FM, $s_i\to(-1)^i s_i$, in Eqn.~\ref{EqnSM16c}, the corresponding free energy $f_{\rm HB}$ per site of long loops (without the last mean field term) can be found within the continuum approximation~\cite{McGurn1975}, valid in the limit $T\ll J$,
\begin{equation}
	\frac{f_{\rm HB}[T,D]}{J} = -\left(\frac{T}{J}\right)^2 \phi\left[\frac{DJ}{T^2}\right] + f_0[T].
\end{equation}
where $-\phi[\rho]$ (with $\rho\equiv DJ/T^{2}$) is the smallest eigenvalue of the quantum-mechanical hindered rotor Hamiltonian:
\begin{equation}
	{\cal H} = -\frac{1}{2}\tilde{L}^{2} + \rho (\cos^2\theta-1/3),
\end{equation} 
$\tilde{L}$ being the angular momentum operator in spherical coordinates.
The $\rho$-independent term $f_0[T]$ is immaterial for the discussion of the phase transition. 

It remains to minimize the mean field free energy per site,
\begin{equation}
	f_{\rm MF} = f_{\rm HB}[D= A T Q_z] + \frac{1}{2} A T Q_z^2
\end{equation}
with respect to the order parameter $Q_z$. Rewriting $T \equiv A |J| \tau$, such that  $D|J|/T^2 = A Q_z|J|/ T = Q_z/\tau$, we have
\begin{equation}
	\frac{f_{\rm MF}}{|J|A^2} = -\tau^2 \phi\left[\frac{Q_z}{\tau}\right]  + \frac{1}{2} \tau Q_z^2. \label{eqnHC1}
\end{equation}
Since $\phi[\rho]\sim\rho^{2}$ for small $\rho$, a local minimum will exist with $Q_{z}=0$. It may be shown that this minimum eventually becomes unstable at increasing temperature, however before this happens, a second minimum at finite $Q_{z}$ gives $f_{\rm MF}=0$ indicating a first order transition.

This happens if, for a positive $Q_z$, one finds simultaneous solutions of $f_{\rm MF}=0$ and  $df_{\rm MF}/dQ_z = 0$, or
\begin{eqnarray}
	\tau^2\phi\left[\frac{Q_z}{\tau}\right] -\frac{1}{2} \tau Q_z^2&=&0,\label{eqnHC2}\\
	\tau\phi'\left[\frac{Q_z}{\tau}\right] -\tau Q_z&=&0. \label{eqnHC3}
\end{eqnarray}
Multiplying the second equation by $Q_z/2$ we find for $\rho= Q_z/\tau$ the equation
\begin{equation}
	\phi[\rho]= \frac{\rho}{2}\phi'[\rho].
\end{equation}
From its solution, $\rho^*$, one obtains the order parameter at the first order transition, 
\begin{equation}
	Q_z^* = 
	\phi'[\rho^*]
\end{equation}
and the transition temperature 
\begin{equation}
	\frac{T_{0}}{A|J|}= \tau^*= \frac{Q_z^*}{\rho^*}.
\end{equation}
Carrying out this procedure numerically, one finds
\begin{eqnarray}
	\rho^*&=& 3.5569,\\
	Q_z^*&=& 0.2377,\\
	\tau^*&=& 0.0668.
\end{eqnarray}
This predicts the equilibrium first order transition to take place at the temperature
\begin{equation}
	\frac{T_{0}}{|J|} = A \tau^* = 0.1 \Delta S.  \label{EqnTPred}
\end{equation}

The mean field prediction for the transition temperature (Eqn.~\ref{EqnTPred}) assumes that all loops contribute to the symmetry breaking field. This is, however, not the case in typical charge-ice samples, since the loops lower than a certain length threshold do not participate in the transition. To a first approximation this can be taken into account by modifying Eqn.~\ref{EqnTPred} to 
\begin{equation}
	\frac{T_{0}}{|J|} = 0.1 c \Delta S=0.05 c \left(\frac{J_{ab}}{J}\right)^{2}, \label{EqnTPred1}
\end{equation}
where $c< 1$ is the average fraction of tetrahedra that are touched by two loops above the length threshold. This fraction will depend on the micro-structure defined via the loop distribution function, an aspect already explored in the previous section giving $c\approx0.97-0.98$ for an $L=8$ charge-ice structure.

\subsubsection{Heat capacity at $T\ll T_{0}$} \label{sm:mft3}

Let us now investigate how the heat capacity behaves for $T<T_{0}$. This is non-trivial due to the temperature dependence of $Q_{z}(T)$. The heat capacity may be evaluated via
\begin{equation}
	c_{V}-c_{V}^{0}=-T\frac{\partial^{2}}{\partial T^{2}}f_{\mathrm{MF}}=-\frac{\tau}{A|J|}\frac{\partial^{2}}{\partial \tau^{2}}f_{\mathrm{MF}},
\end{equation}
where $c_{V}^{0}$ is the heat capacity arising from that part of the free energy not depending on the anisotropy. Using Eqn.~\ref{eqnHC1} together with Eqns.~\ref{eqnHC2} and \ref{eqnHC3}, the above evaluates to
\begin{equation}
	c_{V}-c_{V}^{0}=A\tau\left(-\frac{Q_{z}}{\tau}+5Q_{z}Q'_{z}\right).	\label{eqnHC4}
\end{equation}
The asymptotic form of the low temperature free energy may be found via a quadratic expansion with respect to transverse spin fluctuations around a bulk AFM spin configuration, giving 
\begin{equation}
	\phi[\rho\gg 1]\sim\frac{2}{3}\rho-\sqrt{2\rho}.
\end{equation}

\begin{figure}[h]
	\includegraphics[width=0.7\linewidth,trim=0cm 0cm 0cm 0cm,clip]{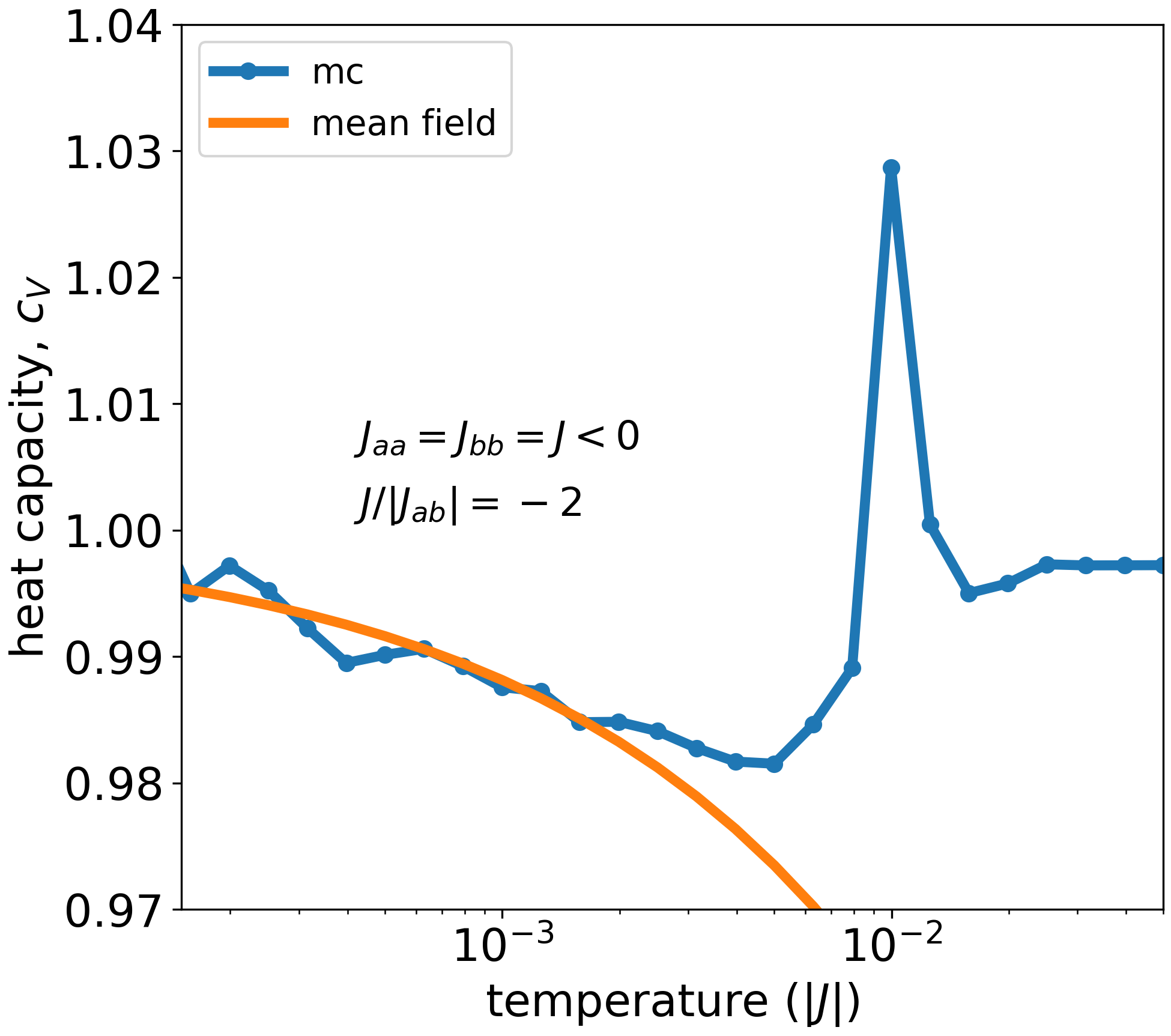}
	\caption{{\bf Low temperature heat capacity asymptote.} Heat capacity as a function temperature for the case of $J/|J_{ab}|=-2$, along with the mean-field prediction for the nematic phase entailed in Eqn.~\ref{eqnHC6}.} \label{figsm6}
\end{figure}

Such an approach entails $c_{V}^{0}=1$, and does not include the effect of $Z_{2}$ domain walls which are exponentially rare and thus contribute only negligibly. Via Eqn.~\ref{eqnHC3}, this yields the leading temperature dependence of the quadrupolar field as
\begin{equation}
	Q_{z}(\tau\ll 1)\approx\frac{2}{3}-\frac{\sqrt{3\tau}}{2}. \label{eqnHC5}
\end{equation}
Substitution of these asymptotics into Eqn.~\ref{eqnHC4} finally gives
\begin{equation}
	c_{V}\sim1-\frac{\sqrt{3}}{2}A\sqrt{\tau}=1-\frac{3}{4}\frac{|J_{ab}|}{|J|}\sqrt{\frac{T}{|J|}}. \label{eqnHC6}
\end{equation}
Thus, in the nematic phase for $T\ll T_{0}$, the heat capacity decreases from unity as the temperature increases. This decrease is due to the softening of the anisotropy (Eqn.~\ref{eqnHC4}) with increasing $T$, rationalizing the simulation result, which also confirms that the effect is proportional to $|J_{ab}/J|$,
see Figs.~1 and ~3 of the main text, and also Figs.~\ref{figsm1} and ~\ref{figsm2}. Fig.~\ref{figsm6} displays the low temperature regime below the transition and the prediction of Eqn.~\ref{eqnHC6}, showing good agreement for the $J/|J_{ab}|=-2$ case, which is within the assumed perturbative regime. The present calculation does not include the temperature dependent features of the heat capacity associated with the release of latent heat close to $T_{0}$, nor does it explicitly take into account that in the limit $T\rightarrow0$, $c_{V}\rightarrow1-\frac{1}{2}N_{\mathrm{loop}}/N$ due to the presence of $2N_{\mathrm{loop}}$ zero modes~\cite{Moessner1998}. It is noted that the factor of one-half originates from the quartic contribution to the heat capacity associated with the emergent anisotropy. For the charge ice structure $N_{\mathrm{loop}}/N\approx0.004-0.005$ -- a number that does not depend strongly on charge-ice realization.

\end{bibunit}
\end{document}